\documentclass[12pt,preprint,epsf]{aastex}
\usepackage{graphicx}% Include figure files
\usepackage{epsfig} %for old style fig. inputs
\slugcomment{Accepted for Publication in The Astrophysical Journal}

\shorttitle{Lyman-alpha Emission in Quasar Absorbers}
\shortauthors{Kulkarni et al.}

\begin{document}


\title{A Fabry-Perot Imaging Search for Lyman-alpha Emission in 
Quasar Absorbers at $z \sim 2.4$}


\author{Varsha P. Kulkarni\altaffilmark{1}}
\affil{Department of Physics and Astronomy, University of South Carolina, 
Columbia, SC 29208}

\author{Bruce E. Woodgate\altaffilmark{1}}
\affil{Code 681, Goddard Space Flight Center, Greenbelt, MD 20771}

\author{Donald G. York\altaffilmark{1,2}}
\affil{Department of Astronomy and Astrophysics, University of Chicago, 
Chicago, IL 60637}

\author{Deepashri G. Thatte and Joseph Meiring}
\affil{Department of Physics and Astronomy, University of South Carolina, 
Columbia, SC 29208}

\author{Povilas Palunas\altaffilmark{1}}
\affil{McDonald Observatory, University of Texas, Austin, TX 78712}

\author{Edward Wassell\altaffilmark{1}}
\affil{St Thomas Aquinas College and the Catholic Univ. of America}


\altaffiltext{1}{Visiting Astronomer, Apache Point Observatory, 3.5-meter 
telescope, owned and operated by the Astrophysical Research Consortium.}
\altaffiltext{2}{Also, the Enrico Fermi Institute}


\begin{abstract}
We have carried out a deep narrow-band imaging survey of six fields with 
heavy-element quasar absorption lines, using the Goddard Fabry-Perot (FP) 
system at the Apache Point Observatory (APO) 3.5-meter telescope. The aim of 
these observations was to search for redshifted Ly-$\alpha$ emission from 
the galaxies underlying the absorbers at $z = 2.3-2.5$ and their companion 
galaxies. The 3 $\sigma$ sensitivity levels ranged between
 $1.9 \times 10^{-17}$ and $5.4 \times 10^{-17}$ erg s$^{-1}$ 
cm$^{-2}$ in observed-frame Ly-$\alpha$ flux. No significant Ly-$\alpha$ 
emitters were detected at $> 3 \sigma$ level. 
The absence of significant Ly-$\alpha$ emission implies limits on the star 
formation rate (SFR) of 0.9-2.7 $M_{\odot}$ yr$^{-1}$ per 2-pixel x 2-pixel 
region, if no dust attenuation is assumed. 
We compare our results with those from other emission-line 
studies of absorber fields and with 
predictions for global average SFR based on the models of cosmic chemical 
evolution. Our limits are among the tightest existing 
constraints on Ly-$\alpha$ emission from galaxies in absorber fields, but are consistent with many 
other studies. In the absence of dust attenuation, these studies 
suggest that SFRs  in a large fraction of objects in the absorber fields may lie below 
the global mean SFR. However, it is possible that dust attenuation 
is responsible for the low emission line fluxes in some objects. It is also possible that 
the star-forming regions are compact and at smaller angular 
  separations from the quasar than the width of our point spread function and, 
  get lost in the quasar emission. 
We outline future observations that could help to distinguish between 
the various possibilities.
\end{abstract}


%% subject header.  Objects should be in the appropriate "individual"
%% headers (e.g. quasars: individual, stars: individual, etc.) with the
%% additional provision that the total number of headers, including each
%% individual object, not exceed six.  The \objectname{} macro, and its
%% alias \object{}, is used to mark each object.  The macro takes the object
%% name as its primary argument.  This name will appear in the paper
%% and serve as the link's anchor in the electronic edition if the name
%% is recognized by the data centers.  The macro also takes an optional
%% argument in parentheses in cases where the data center identification
%% differs from what is to be printed in the paper.



\section{INTRODUCTION}

A great deal of progress has been made in the last decade in the studies of 
distant galaxies. The average star formation history of the 
Universe has also been estimated from emission properties 
of galaxies detected in deep imaging and redshift surveys such as 
the Canada-France Redshift Survey and the Hubble Deep Field (e.g., 
Lilly et al. 1996; Madau et al. 1996, 1998). A large population of bright 
galaxies with high star formation rates (SFRs) has been uncovered by 
means of the Lyman-break 
technique (e.g., Steidel et al. 1999). However, such flux-limited 
studies may not be adequate for investigating the evolution of normal  
galaxies, since such studies are biased toward the brighter or the more 
actively star forming galaxies. In principle, a less biased sampling of 
distant galaxies 
may be obtained by means of the absorption lines they superpose in the 
radiation from background quasars. 

The presence of heavy-element absorption lines in quasar spectra preselects 
regions that have undergone some degree of star formation. There is 
nothing 
special about the lines of sight to the absorbers except for the 
convenient placement of background quasars, assuming that most of the 
absorbers are 
  intergalactic material and not ejected material near the QSO at 
  high, special relativistic velocities. Thus, it is very likely that these absorbers are surrounded 
by other galactic or protogalactic objects at the same redshifts. Thus in 
principle, deep images of quasar absorber fields may 
reveal star-forming regions at high redshift. 

Since a well-known signature of star forming galaxies is strong Ly-$\alpha$ 
or H-$\alpha$ emission, a natural strategy to look for the absorber galaxies 
or their companions is to search for the redshifted 
emission lines. 
Several previous studies have attempted to detect continuum and line 
emission from galaxies underlying quasar absorbers. At low redshifts,  
[O II] or [O III] emission has been 
detected in Mg II systems with narrow-band imaging (e.g., 
Yanny 1990; Yanny, York, \& Williams 1990; Bergeron \& Boisse 1991; Yanny \& York 1992). 
Searches for low-$z$ damped Lyman-alpha absorbers (DLAs) have often imaged, 
and sometimes spectroscopically 
confirmed, galaxies with various morphologies (spirals, irregulars, low 
surface brightness galaxies etc.--e.g., Steidel et al. 1994, 1995; 
LeBrun et al. 1997; Bowen et al. 2001; Cohen 2001; Turnshek et al. 2001). 
At high redshifts, however, it has been much more difficult to identify 
and study in detail the galaxies responsible for quasar absorption systems. 
There have been a few detections of Ly-$\alpha$ emission in quasar 
absorber fields (e.g. Lowenthal et al. 1991; Francis et al. 1996; 
Roche, Lowenthal, \& Woodgate 2000). However, 
most other attempts to detect Ly-$\alpha$ emission from high-$z$  
intervening ($z_{abs} < z_{em}$) DLAs have produced either 
non-detections or weak detections (e.g. Smith et al. 1989; Hunstead et al. 
1990; Lowenthal et al. 1995; Djorgovski et al. 1996). 
Most of the few confirmed Ly-$\alpha$ 
detections for high-redshift DLAs have been 
for absorbers with $z_{abs} \approx z_{em}$, which may differ from the 
cosmologically more interesting general population of DLAs with 
$z_{abs} < z_{em}$ (e.g. Warren \& Moller 1996; 
Moller \& Warren 1998; Fynbo et al. 1999). Most attempts to 
detect DLAs in H-$\alpha$ have either yielded non-detections or detected 
objects 
separated by large angular distances from the quasars, rather than 
objects close to quasar sightlines (see, e.g., Teplitz, Malkan, \& McLean 1998; Bechtold et al. 
1998; Mannucci et al. 1998, Bunker et al. 
1999). These objects, while unlikely to be the absorbing galaxies themselves, 
are still interesting because they are likely to be companion galaxies in the same cluster 
as the absorber. 
A summary of previous attempts to detect high-$z$ DLAs in 
emission is given by Kulkarni et al. (2000). 

With the goal of obtaining a large sample of SFR estimates in the fields 
of high-redshift absorbers, we have started a Lyman-$\alpha$ imaging 
survey for absorbers at $2.3 < z < 2.6$ using a Fabry-Perot imager as a 
tunable narrow-band filter. Here we report our results for six fields with 
known heavy-element absorption systems. Sample selection, observations, 
and data reduction are described in section 2. The reduced images and photometry 
are presented in section 3. Section 4 presents a comparison of our results 
with previous studies of the same fields and with results of other Ly-$\alpha$ 
emitter studies. Section 4 also compares the SFR constraints 
from our study with the estimates from other emission-line 
searches for quasar absorber galaxies. A comparison with predictions based 
on global SFR models is also presented. 

\section{OBSERVATIONS AND DATA REDUCTION}

\subsection{Sample Selection}

The blue and vis-broad etalons in the Goddard Space Flight Center (GSFC) Fabry 
Perot (FP) imaging system have optimum sensitivity and resolution in the 
wavelength range $\sim 4000-5000$ {\AA}. We 
therefore restricted our search to the redshift range $2.3 < z < 3.1$. 
We searched the York et al. (1991) catalog of heavy-element quasar absorbers 
for absorbers with (i) $2.3 < z_{abs} < 3.1$  (ii) $z_{abs} < z_{em} - 0.6$ 
to avoid absorbers possibly associated with the quasars, and (iii) with 
well-detected mixed 
ionization species (Si II, Al II or O I in addition to C IV and/or Si IV). Six 
such objects were finally observed. Two of these six systems are DLAs, and  
one is a sub-DLA. 
In addition to these six systems, we also observed the  well-studied field 
of the radio galaxy 53W002 containing Ly-$\alpha$ emitters 
at $z \approx 2.4$ (Pascarelle et al. 1996a, 1996b; Keel et al. 1999), as a 
calibration object. 
Table 1 lists the general properties of the objects observed.

\subsection{Observations}

The FP instrument used for our observations has a choice of six Queensgate 50 mm diameter 
piezo-electrically driven, capacitance-stabilized etalons which can be 
tuned to any wavelength in the range 4000-10000 {\AA}, and with resolution from 
2.5 {\AA} to 28 {\AA} FWHM depending on the etalon and the wavelength. 
The system throughput 
including the CCD is 20\% in the red and 5\% in the blue. The instrument is 
used at the Nasmyth f/10.3 focus of the APO 3.5 m telescope. Behind the 
telescope focus a field lens and collimator lens collimate the light through 
the etalon and an order-sorting blocking filter, and a camera lens refocuses 
onto the CCD.

The camera has a STIS technology SITe 2048x2048 CCD with 21 $\mu$m pixels, 
5 e$^{-}$ rms readout noise, MPP (inverted operation), minichannels, 
and a very high efficiency down to 3400 {\AA}. The quantum efficiency is 82\% at 4000 {\AA}, 90\% at 7000 {\AA}  
and 53\% at 9000 {\AA}. This provides a large field of view, low noise, smooth 
bias levels, high charge transfer efficiency, and high sensitivity with UV, 
visible and IR etalons. The instrument can be controlled from the observatory 
control room via the Apache Point fiber optics communication system, or 
remotely over the internet, using command line control and a VNC GUI 
monitoring system. 
        
The wavelength transmitted by an FP system is a function of the field
angle. $\Delta \lambda= \lambda [1/(1+(r/f)^{2})]$, where $f$ is the camera lens 
focal length, 
 200 mm, and $r$ is the distance on the detector from the optical axis. 
 At the APO f/10.3 Nasmyth focus, $\Delta \lambda \approx \lambda 
\biggl\{1- [1/\{1+(5.8 \times 10^{-4} \theta)^{2}\}] \biggr\}$, with 
 $\theta$ in arcseconds  
 on the sky. For example, for $\lambda$ = 4100 {\AA}, $\theta = 90$ \arcsec, 
$\Delta \lambda$ = 11.1 
 {\AA}. The effective monochromatic field of view depends on the tolerable 
 fraction of the spectral width.
 
Table 2 lists the details of our observations. 
The observations were carried out primarily with the blue etalon 
during 9 runs between October 2000 and 
May 2004. 
The seeing varied between 1.0$\arcsec$ and 1.8$\arcsec$ FWHM. Conditions were 
not photometric for many of the nights. The narrow-band exposures were obtained 
setting the FP to the air wavelengths of the expected Ly-$\alpha$ emission 
lines at the absorber redshifts. In some cases, additional narrow-band exposures 
were also obtained with the vis-broad etalon at wavelength settings offset by 
400 km s$^{-1}$ redward of the absorber 
redshifts.  The wavelength calibration was checked throughout using scans of 
lines from an Ar lamp. A temperature control program was run throughout 
to automatically keep the FP wavelength setting fixed by correcting for 
effects of temperature variations every five minutes. The FWHMs for the different fields were 5.6-14.7 {\AA}, as measured from the width of the calibration 
lamp lines. A $2 \times 2$ binning was used, resulting in a plate scale of 0.366 $\arcsec$ 
per binned pixel. The full circular field of view corresponds to 
$\sim 3.4 \arcmin$ in diameter. Of course, given the variation in the central 
wavelength as a function of angle from the axis, the central 
$\approx 2.1 \arcmin$ diameter region is sampled within $\pm 400$ km s$^{-1}$ of the 
central wavelength expected at the absorber redshift at 4100 {\AA}.

Total narrow-band integration times were 
320-600 minutes per field, split into 8-15 exposures of 40 minutes each. 
In addition, B-band exposures were obtained for all fields to sample the 
UV continuum near the Ly-$\alpha$ line. 
A pentagonal dither pattern was used to aid with flat-fielding and cosmic 
ray removal.  Twilight sky flats 
were obtained for each wavelength setting. Standard star exposures were 
obtained for HZ44. Additionally, 
images of the 53W002 field 
were also used for photometric calibration. 

\subsection{Data reduction}

Data reduction was carried out using standard IRAF tasks. 
The biases on each night were combined to get an 
average bias frame. The darks, flats and the object frames were corrected for 
the overscan region using the ccdproc task. All the images were trimmed 
to a size of 561 x 561 pixels or $3.4 \arcmin \times 3.4 \arcmin$, which 
contains the full field of view of the Fabry Perot. The darks, the flats  
and the object frames on each night were bias-corrected with the average bias  
corresponding to that night. The flat frames and the object frames were then 
dark-corrected with the combined dark for each night. The flat frames were combined 
using the flatcombine task to produce average flats corresponding to 
different wavelengths and filter settings. In some cases, twilight or dome 
flats were not available for a particular wavelength and filter setting. 
In such cases, dark-corrected 
object frames were combined (with all objects removed) to create an 
average sky flat that was later 
used for flatfielding. Object frames for a particular wavelength and 
filter settings were flat fielded using the average flats in the same setting. 
Bias, dark and 
flat frames from other nights in the same season were used in a very few cases 
where they were not available on the same night. The IRAF task cosmicrays was 
used to remove cosmic rays from each object frame. A bad pixel mask was 
created for each image and the bad pixels were removed using the IRAF tasks 
ccdproc or fixpix.

To account for the varying degrees of extinction encountered over different 
nights, we carried out an empirical correction. 
The flux counts for unsaturated bright stars were measured in each 
image using the phot task in IRAF. The counts were corrected for 
images with different exposure times. For each star, the maximum 
value of counts $f_{max}^{j}$ from different images was taken as an 
indicator of the unextinguished flux level of the $j$th star. The flux 
ratio for the $j$th star in the $i$th image $f_{max}^{j} /f_{i}^{j}$ 
was then calculated for the best four or five stars in all images 
to estimate the extinction factors. 
The extinction factors thus derived for these four or five stars in each given  
image were then averaged to get an indication of the extinction for 
that image. Finally, each image was multiplied by this average empirical 
extinction factor to get an extinction corrected image.

All extinction-corrected images for each field and filter/wavelength setting 
were registered and shifted so as to match the coordinates 
of a reference star in all frames. This was done using the imexam, 
lintran and imshift tasks. Some images were rotated using the geomap 
and gregister tasks to match the coordinates of all stars in different 
images, to correct for 
a small rotational offset between some images taken at the same pointing in 
different observing seasons. Finally, the individual exposures for a given object and given 
wavelength and filter setting were combined using the IRAF task imcombine 
to get the final narrow-band and broad-band images.

\section{RESULTS}

Figures 1(a)-6(a) show $2.1 \arcmin \times 2.1 \arcmin$  
sections of the reduced broad-band (B) 
images of the quasar absorber fields 
of Q0216+080, Q0636+680, Q0956+123, Q1209+093, Q1442+101,  
and Q2233+131, 
respectively. The corresponding panels (b) show the narrow-band images of the same 
fields obtained with the blue etalon. 
The stripes at the borders of some images are an artifact of the coadding of 
the dithered images. The quasar is absent or considerably dim in the 
narrow-band images for Q0216+080, Q1209+093, and Q2233+131 since the absorbers being 
studied in these systems are DLAs or sub-DLAs. Figure 7 shows the calibration field 53w002. 
 
The expected B-band continuum must be subtracted from the observed narrow-band 
fluxes in order to determine if a statistically significant redshifted Ly-$\alpha$ 
excess exists for any object seen in both bands. We estimated the continuum 
in the narrow-band images by scaling the B images using the relative 
photometric calibrations of the two images. To do this, we subtracted a 
scaled B band image from the narrow-band image, aligning the images spatially, 
and adjusting the scaling factor so as to minimize variance in the central 
portion of the subtracted image. Furthermore, to minimize the effect of 
different seeing in the broad-band and narrrow-band images, the broad and 
narrow-band images were convolved with a Gaussian of the same FWHM before subtraction in IDP3. 
The subtraction was carried out interactively, using the interactive data language 
(IDL, version 6.1) program Image Display Paradigm-3 (IDP-3) version 2.7, 
written by D. Lytle and E. Stobie (see, e.g., Lytle et al. 1999)
Figs. 1(c)-7(c) show the resultant continuum-subtracted 
images for the six 
quasar absorber fields and the calibration field of 53w002. Finally, Figs. 1(d)-7(d) 
show $1 \arcmin \times 1 \arcmin$ close-up views of the continuum-subtracted images. 

All of the objects in the narrow-band images disappeared almost completely 
after subtraction of the continuum. The slight residuals left 
at the positions of some objects arise because of the difficulty in 
matching the point spread functions (PSFs) perfectly in the broad and narrow-band images. No significant Ly-$\alpha$ emission at $> 3 \sigma$ level was detected from any 
object in any of the quasar absorber fields. The few vis-B images obtained for three of the fields 
(Q0216+080, Q0636+680, Q2233+131) were also analyzed in a similar manner, 
and showed no Ly-$\alpha$ detections either. No significant objects 
were found even if the images were smoothed using Gaussian filters 
of about 2.5 or 5 pixels FWHM. Finally, we also compared our images with 
broad-band images from the Sloan Digital Sky Survey (SDSS) where available. A few 
very faint features are 
seen in our images for Q0216+080, but these features are indistinguishable from the 
noise. In Q0636+680 and Q0956+123, some residual flux is seen 
close to the quasar after continuum subtraction. However, these regions 
close to the quasars are not well sampled at our modest 
angular resolution. Thus, it is not clear whether the 
residuals seen near Q0636+680 and Q0956+123 are significant features or artifacts, without obtaining higher 
resolution images of these fields. In the case of Q0956+123, the feature apparent near the left upper corner of 
Fig. 3(d) is an artifact arising from a residual multipixel cosmic ray event in a 
single narrow-band image. In the field of Q2233+131, three 
bright galaxies are seen:  
17.4$\arcsec$ east and 1.7$\arcsec$ south of the quasar; 28.2$\arcsec$ east 
and 51.7$\arcsec$ south of the quasar; and 35.2$\arcsec$ east, 
30.2$\arcsec$ north of the quasar (Fig. 6c). All of these galaxies are 
seen in the SDSS. However, all of these galaxies are  
far too bright to be at $z=2.55$ ($g=19.76, 20.35, 20.49$ respectively 
from SDSS). It is more likely that the 
excess emission seen in these objects is [O II] $\lambda 3727$ emission 
from interloper galaxies at $z=0.16$. We plan to obtain spectra of these  
galaxies in the near future to determine their redshifts. 

To estimate the limits on the Ly-$\alpha$ fluxes for 
the absorbers in our fields, we used our calibration observations of the 
field of the radio galaxy 53w002. This field is known to have Ly-$\alpha$ 
emitters at $z = 2.39$ (Pascarelle et al. 1996a, 1996b; Keel et al. 1999). 
For calibration purposes, we used the brightest objects seen in our 
continuum-subtracted narrow-band images of this field, i.e., objects 3 and 1 of Keel et al. 
(1999; table 2), labeled 
O1 and O2 in the bottom panel of Figure 7. Aperture photometry of these two objects was carried out using the IRAF task phot with aperture sizes $\sim 3$ 
times the seeing FWHM. Using the Ly-$\alpha$ 
fluxes for these from Table 2 of Keel et al. (1999) and the counts 
measured for those objects in our images, we estimated the 
photometric calibration of our images. To correct for sensitivity differences 
at the different wavelengths used for our quasar absorbers and the calibration 
field 53w002, we used our observations of standard star HZ44. Using the photometry of this star in images obtained at the different wavelength settings, 
and adopting the absolute spectral distribution of this star from Oke (1990), 
we estimated the relative sensitivity differences of the FP at the different wavelengths. 

We next estimated the noise level in the continuum-subtracted blue-etalon 
images for 
each field. To do this, we considered a $10 \arcsec \times 
10 \arcsec$ region centered on the quasar and determined the noise 
profile over this region using IDP-3. While doing this, we disregarded a circular 
region centered on the quasar with a radius of roughly twice the seeing FWHM to 
avoid effects of imperfect seeing matches between the broad and narrow band 
images. We then measured the mean noise level per pixel in concentric annuli 
of radii ranging from 2 times the seeing to 5 $\arcsec$, and finally corrected for the sensitivity differences at the various wavelengths. The noise level per 
pixel was found to be nearly constant everywhere (within a few $ \%$)  
in this region. The mean value of the noise per pixel was used to estimate 
the noise level per 2 pixel x 2 pixel region, assuming Poisson statistics. 
Based on this,  we 
estimated the 3 $\sigma$ observed-frame Ly-$\alpha$ point source flux sensitivity reached 
in our images. If dust attenuation is assumed to be small, these Ly-$\alpha$ flux 
limits can be converted to limits on the star formation rates (SFRs). These point-source SFR 
limits implied by the non-detections of 
Ly-$\alpha$ emission in our quasar absorber fields are listed in Table 3 and are in the 
range of 0.9-2.7 M$_{\odot}$ yr$^{-1}$. Here, we have used 
the prescription of Kennicutt (1998) for converting H-$\alpha$ luminosity 
$L_{H-\alpha}$ to the 
SFR, i.e., $SFR (M_{\odot} \, yr^{-1})= 7.9 \times 10^{-42} L(H-\alpha) (erg \, s^{-1})$ 
and assumed a ratio $L_{Ly-\alpha}/L_{H-\alpha} = 8.7$ for case-B 
recombination. 
 
We also used our continuum-subtracted images to estimate the flux limits 
for a diffuse foreground absorbing galaxy. To do this, we used 
the mean value of the noise per pixel to estimate 
the flux limit over the entire 10 $\arcsec \times 10 \arcsec$ region. This 
flux limit was used to calculate the limiting Ly-$\alpha$ luminosity and 
hence the limiting SFR in the $10 \arcsec \times 10 \arcsec$ region. At $z =2.4$, 
this region corresponds to a size of $\approx 81 \times 81$ kpc$^{2}$ centered on the 
quasar, for the comsology adopted here. The diffuse-source flux 
limits are in the range of 2.6-7.4 $\times 10^{-16}$ erg s$^{-1}$ cm$^{-2}$ summed 
over the $10 \arcsec \times 10 \arcsec$ region. The corresponding summed SFR limits 
are in the range of 12.3-36.4 M$_{\odot}$ yr$^{-1}$ over the 
entire $10 \arcsec \times 10 \arcsec$ region.

\section{DISCUSSION}

\subsection{Comparison with Other Imaging Studies of Our Targets}

For most of our targets, no other imaging information exists in any waveband 
on a scale comparable to our fields of view. For Q0216+080 and Q2233+131, 
HST NICMOS H-band images are available (Warren et al. 2001). 
The NICMOS images of Q0216+080 show two objects with $H_{AB} = 25.75$ and 24.34, 
located  
1.4$\arcsec$ and 3.8$\arcsec$ away from the quasar, at position angles of 
-104.8 and 97.8 degrees east of north, respectively. However, the closer 
object is 
likely to be an artifact since it occurs on a diffraction spike. For 
Q2233+131, the NICMOS H-band image revealed two objects with $H_{AB} = 25.05$ 
and 25.12, located 
2.8$\arcsec$ and 3.3$\arcsec$ away from the quasar, at position angles of 
158.6 and 68.3 degrees east of north, respectively. Of these, the former 
object has been identified with a Lyman limit system at $z=3.15$ (Djorgovski et al. 
1996). None of these objects are seen in our B-band or NB images. This could 
be because of the difference in angular resolutions of 
our study and the NICMOS study. In either case, 
no information on star-forming emission-line objects is available from these broad-band 
images. The  only additional emission-line constraints available are for 
Q0216+080 and Q2233+131. The narrow-band imaging study of Deharveng et al. (1990) 
found no Ly-$\alpha$ emitters in the Q0216+080 field, placing a 3 $\sigma$ upper 
limit of $6.9 \times 10^{-16}$ erg s$^{-1}$ cm$^{-2}$ 
on the Ly-$\alpha$ emission flux. Our observations of this field 
have provided a Ly-$\alpha$ flux limit $> 10$ times tighter than that of 
Deharveng et al. (1990). For Q2233+131, our limit agrees closely (within 
$\approx 20 \%$) with that of Lowenthal et al. (1995). 

\subsection{The Space Density of Ly-$\alpha$ Emitters: Comparison with Other 
Studies} 

To understand whether the non-detections of Ly-$\alpha$ emitters (LAEs) 
in our fields 
are consistent with other surveys for Ly-$\alpha$ emitters, we now examine 
results from some recent LAE searches. Stiavelli et al. (2001) detected 
58 LAE candidates over a field of 1200 arcmin$^{2}$ at 
$z=2.422 \pm 0.072$. Based on this, they deduced a completeness-corrected 
space density of 0.07 LAEs per arcmin$^{2}$ with Ly-$\alpha$ fluxes above 
$2 \times 10^{-16}$ erg cm$^{-2}$ s$^{-1}$. Palunas et al. (2004) detected 
37 absorbers in a 46' $\times$ 46' field of view. Based on this, they derived the 
space density of LAEs to be 0.019 arcmin$^{-2}$ in a redshift interval 
$\Delta z  = 0.045$ at $z \sim 2.4$. These (Stiavelli et al. 2001, Palunas 
et al. 2004) studies imply 0.49 and 0.42 LAEs per arcmin$^{2}$ per unit 
redshift, respectively. At $z = 4.5$, 
LAE searches by Rhoads et al. (2000) and Malhotra \& Rhoads (2002) have 
found space densities of 4000 deg$^{-2}$ per unit redshift, i.e. 1.1 LAE 
per arcmin$^{2}$ per unit redshift. 

These other LAE searches cover much wider fields of view, and larger redshift 
ranges $\Delta z$ (wider filter bandwidths) than our study. Adopting a mean 
space density of 0.45 LAEs per 
arcmin$^{2}$ per unit redshift at $z=2.4$ from Stiavelli et al. (2001) 
and Palunas et al. (2004), and using $\Delta z \approx 0.003$ as a typical 
redshift range covered by our study, one would expect  $< 1 $
absorber in our fields with effective monochromatic 
coverage of $\approx 2.1'$ , if our fields were 
similar to the other LAE fields. 

On the other hand, the LAE searches of Rhoads et al. (2000), Stiavelli et al. (2001), 
and Malhotra \& Rhoads (2002) are not selected by the presence of a quasar absorption system. 
The presence of a well-established heavy-element quasar absorber with mixed ionization 
implies the  existence of a region that has had some star formation. 
At $z \sim 2.4$, our complete 
fields of view of about 3.4' diameter cover about 2.2 Mpc$^{2}$ around the 
absorbing 
sightline (assuming $\Omega_{m} = 0.3$, $\Omega_{\Lambda} = 0.7$, and $H_{0} = 70$ km s$^{-1}$ 
Mpc$^{-1}$). Given the gradient in the central wavelength, the 
roughly monochromatic (within $\pm 400$ km s$^{-1}$ of the absorber) fields of view of 
about 2.1' diameter cover about 0.8 Mpc$^{2}$ around the quasar sightline.  
Since the absorbing galaxy certainly must lie within this region, 
one might expect a higher LAE density in our fields than that in a 
blind field. Francis et al. (1996, 2004) and Palunas et al. (2004) did indeed find a number of 
LAEs in wide-field surveys surrounding a field with 3 sub-DLAs. However, this 
region appears to be a filament with higher density than a typical field region. 
Furthermore, the LAEs in this filament are separated from 2 of the 3 absorber 
sightlines by more than 2' in radius (see Fig. 4 of Palunas et al. 2004).  
Such objects would thus not be detectable in a field of view such as ours. Thus, overall, the lack of LAEs in our observations are 
not inconsistent with the space density of LAEs found by other studies. We note,  
however, that detailed comparisons would require a full consideration of the 
luminosity function of LAEs since our observations reach fainter flux limits than 
many of the other LAE searches.

\subsection{The Star Formation Rates of Quasar Absorbers and their Companions}

Fig. 8 plots our SFR limits in the fields of quasar absorbers 
together with the results of other searches 
for Ly-$\alpha$, H-$\alpha$, H-$\beta$, [O II], and [O III] emission 
in quasar absorber fields. The 
filled red triangles show our APO Ly-$\alpha$ limits, while 
the unfilled black triangles at $z \approx 1.9$ show our limits from previous 
H-$\alpha$ imaging with HST/NICMOS (Kulkarni et al. 2000, 2001). Other data are from 
Yanny et al. (1987); Yanny, York, \& Williams (1990); Yanny (1990); Hunstead 
et al. (1990); Deharveng et al. (1990, 1995); Giavalisco et al. (1994); 
Lowenthal et al. (1995); Francis et al. (1996); Warren \& M\'oller (1996);  
Bergvall et al. (1997); Mannucci et al. 
(1998); Teplitz et al. (1998); Bunker et al. (1999); Fynbo et al. (1999, 2000); 
Bouche et al. (2001); van der Werf et al. (2000);  
M\'oller et al. (2002); Meyer, Thompson, \& Mannucci (2003); M\'oller, Fynbo, \& Fall (2004); 
Christensen et al. (2004, 2005); 
Schulte-Ladbeck et al. (2004); Chen, Kennicutt, \& Rauch (2005); 
and Weatherley et al. (2005). In total, we have plotted 71 detections and 30 
upper limits in Fig. 8. We note, however, that many of the 
emission line detections are for candidates that 
have not yet been confirmed 
with spectroscopy or multiple narrow-band imaging. For the few objects 
where multiple emission lines have been detected, we have plotted the 
values from H-$\alpha$, H-$\beta$, or 
[O II], in that order of preference. All the shown candidate 
detections and 3 $\sigma$ upper limits have been normalized 
to $\Omega_{m} = 0.3$, $\Omega_{\Lambda} = 0.7$,  and $H_{0} = 70$ km s$^{-1}$ Mpc$^{-1}$. Once again, 
the prescription of Kennicutt (1998) has been used for the conversion 
from the H-$\alpha$ luminosity $L_{H-\alpha}$ to the 
SFR, and a ratio $L_{Ly-\alpha}/L_{H-\alpha} = 8.7$ for case-B 
recombination has been assumed. No correction has been made for dust attenuation, but 
we examine this issue in more detail below. Our APO FP constraints are among the 
lowest, but 
are clearly consistent with many other measurements. The open inverted triangles 
at $0 < z < 2.5$ are based on spectroscopic (slit) observations of H-$\alpha$. 
The unfilled circles are candidate objects with H-$\alpha$ emission detected at the 
absorber redshift, but are often more than $\sim 10 \arcsec$ away from the 
quasar, and have a lower precision in redshift. 

The curves in Fig. 8 are useful for comparing the 
data with global predictions based on the luminosity density of 
galaxies from the deep galaxy 
imaging surveys. Since most of the data in Fig. 8 are for objects in 
DLA fields, we show the 
calculations of Bunker et al. (1999) for the predicted cross-section-weighted SFR in DLAs. 
These models use two alternative sets of assumptions about the number 
of absorbers per unit redshift and the distribution of global SFR among 
individual absorbers at a given redshift.
The thick and thin solid curves show the LD5 and LD0 ``large disk'' calculations  
(for $q_{0} = 0.5$ and $q_{0} = 0$ respectively), of Bunker et al. 
based on the closed-box Pei \& Fall (1995) models. The LD models assume that DLAs are the 
progenitors of spiral galaxies, with space density equal to that of local 
spirals, but with size and SFR of each DLA larger in the past compared to that 
of a local spiral. 
The dashed curve shows the H5 and H0 predictions of Bunker et al. (1999) 
(for $q_{0} = 0.5$ and $q_{0} = 0$ respectively) for the ``hierarchical'' 
hypothesis, which assumes that there were multiple DLAs at high 
redshift corresponding to every present-day spiral (i.e. that DLAs were 
sub-galactic fragments that later merged to form present-day spirals). 
Thus in the H5 and H0 models, DLAs have 
the same distribution of gas cross-section sizes as in local spirals, but with 
a higher space density in the past, and have smaller SFRs than 
in the corresponding ``large-disk'' models. 

The LD curve, if computed for $\Omega_{m} = 0.3$, $\Omega_{\Lambda} = 0.7$,  
and $H_{0} = 70$ km s$^{-1}$ Mpc$^{-1}$, would lie between 
the LD5 and LD0 curves, since the distance and time scales for the former 
cosmology are intermediate between those for $q0=0$ and $q0=0.5$. 
This can also be seen by comparing the comoving global SFR  
predicted by 
the Pei \& Fall (1995) closed-box models for $q_{0} = 0.5$ and $q_{0} = 0$, with that 
predicted by the $\Lambda$-cold dark matter ($\Lambda$-CDM) 
hydrodynamical simulations of Nagamine, 
Cen, \& Ostriker (2000). 
Thus, a large fraction of the observed SFRs would appear to fall significantly below the 
prediction of the large-disk scenario. The H curve for this $\Lambda$-CDM cosmology would 
be identical to the H5 and H0 curves (since the latter curves coincide with each other). 
A large fraction of the data points in Fig. 8 thus appear to be more 
consistent with the hierarchical scenario, suggesting perhaps that many 
absorbers or their companion galaxies may be arising in star-forming dwarf galaxies or sub-galactic fragments that merged later 
to form present-day galaxies. However, several detections and 
upper limits, including our FP limits, appear to be considerably below even the 
hierarchical prediction. Indeed, about 63$\%$ of the detections and about 73$\%$ of the 
limits plotted in Fig. 8 
are less than 5 $M_{\odot}$ yr$^{-1}$. Such large local deviations for the absorber fields 
from the global mean would be surprising. 

Taken at face value, Fig. 8 suggests that the absorption-based view of the cosmic star 
formation history could be quite different from the emission-based view, i.e., 
from the star formation history inferred on the basis of the direct 
galaxy imaging surveys such as the Hubble Deep Field. Low SFRs for 
galaxies in absorber fields would also be consistent with 
the low global metallicities found in DLAs (e.g., Kulkarni et al. 2005 and 
references therein). Similar suggestions have also been made by Wolfe et al. (2003), who inferred 
relatively high SFR values for some DLAs based on C II* absorption, but found the high SFRs to be inconsistent 
with the low metallicities observed in the DLAs. We note, however, that 
while the C II* method may potentially offer 
an interesting way to estimate SFRs in DLAs, it seems less direct 
and more model-dependent than the emission-line based constraints. 

There are also other possible interpretations of Fig 8. It is possible that 
the true SFRs in galaxies in the absorber fields are higher, but their emission 
lines, especially Ly-$\alpha$, 
are attenuated by dust. It is well known that resonance scattering of 
Ly-$\alpha$, in the presence of dust, can lead to quenching of the 
Lyman-$\alpha$ emission in high-$N_{\rm H I}$ systems 
  (e.g., Charlot \& Fall 1991). Presence of dust in some Ly-$\alpha$ 
emitting regions is also suggested from the recent detection of 24 $\mu$m 
emission in Spitzer Space Telescope images of LAEs at $z=2.4$ in an absorber  
field (Colbert et al. 2004). On the other hand, there are several reasons to 
expect that Ly-$\alpha$ emission 
may still be seen from absorbers and their companion galaxies, in at least some cases. First, the very 
knowledge of a DLA means that the quasar in which it appears is not greatly  
affected by dust attenuation. Ly-$\alpha$ is easily seen from regions of 
such low extinction. Second, the dust-to-gas ratios in DLAs inferred 
from relative abundances such as [Cr/Zn] or [Fe/Zn] are much smaller 
than in the Milky Way (e.g., Pettini et al. 1997; Khare et al. 2004; and references therein). 
In a recent study of a large number of SDSS quasar spectra, we have recently 
found evidence for a statistically significant but small amount of dust in 
quasar absorbers (York et al. 2005). Finally, there are several examples 
of high-$z$ Ly-$\alpha$ emitters with very little dust attenuation 
(e.g., Giavalisco et al. 1994; 
Francis et al. 1996; Hu, Cowie, \& McMahon 1998; Kudritzki et al. 2000). 

The effect of dust attenuation is expected to be much less severe in the 
H-$\alpha$ line. Indeed, the SFR constraints based on H-$\alpha$ detections in most high-$z$ 
DLA candidates or companions shown in Fig 8 are considerably higher than 
the constraints 
from the Ly-$\alpha$ emission searches.  Considering detections alone, 
the median SFRs from H-$\alpha$, Ly-$\alpha$,  
and [OII] are 35.9, 3.2, and 0.3 $M_{\odot}$ yr$^{-1}$, 
respectively. If both detections and limits are considered, the median SFRs 
from H-$\alpha$, Ly-$\alpha$, 
and [OII] are 28.2, 2.4, and 0.3 $M_{\odot}$ yr$^{-1}$ treating the limits 
as detections, and 18.8, 0.1, and 0.3 $M_{\odot}$ yr$^{-1}$ treating the limits 
as zeros. It is tempting to think that the difference between the median H-$\alpha$ and Ly-$\alpha$ values 
could be partly caused by dust extinction. However, it is not 
clear whether this is a significant effect, since Ly-$\alpha$ measurements (candidates or limits) 
are not available for most of the H-$\alpha$ candidates. In the few objects 
that show detections of 
multiple emission lines, the SFR estimates from the available lines are 
usually consistent within a factor of $\sim 2$. Also, some of 
the H-$\alpha$ candidates are at large angular separations from the quasars 
and have less accurate redshifts. Thus, there is no clear indication that Ly-$\alpha$ in the 
H-$\alpha$ candidates is attenuated by dust. In any case, dust alone 
may not explain low Ly-$\alpha$ 
emission in every individual case, because the extent of Ly-$\alpha$ attenuation 
also depends on several other factors, such as the orientation, 
the topology of the H II region, and the distribution of 
stars, gas and dust within the absorber. Finally, dust attenuation cannot 
explain why nearly all of the SFRs inferred from [O II] and $\approx 25 \%$ of 
the SFRs inferred from H-$\alpha$ are low. 

Another possibility is that the star-forming regions in the absorber galaxies 
are compact, lie directly in front of the quasars, and hence get lost in 
the quasar point spread function (PSF) in our study and other ground-based 
studies so far. For example, if the absorbing galaxies had star-forming cores 
comparable in size to those in the Lyman-break galaxies [typically $\sim 1.6 \, 
(H_{0}/70)^{-1}$ kpc]  
aligned perfectly in front of the quasars, it would not be possible to 
resolve them with studies such as 
our own (and most other ground-based studies used in Fig. 8). In other words, 
the SFR constraints derived from such studies may be more appropriate for the companion galaxies of the absorbers, 
rather than the absorbers themselves. It is difficult to rule out this possibility without  
high-resolution imaging data. However, the chance of having 
the star forming region exactly in front of the quasar in {\it each} case 
is small. Indeed, the absence of emission within the Ly-$\alpha$ 
absorption profiles in the published 
spectra for most DLAs (including those from our study) suggests that they 
do not have star-forming regions perfectly aligned with the background 
quasars. Even if the absorber galaxies themselves were small, 
they are unlikely to be isolated objects, since such small galaxies are 
expected to occur as companions to larger galaxies within a few hundred kpc. 
In any case, it would still be surprising why the few existing high-resolution 
imaging studies of absorber galaxies have often failed to detect H-$\alpha$ emission 
or found relatively weak Ly-$\alpha$ emission (e.g., Bouche et al. 2001; 
Kulkarni et al. 2000, 2001; M\'oller et al. 2002). 

Future narrow-band imaging studies of more high-redshift quasar absorber fields 
would help 
to understand whether the low Ly-$\alpha$ fluxes we find are representative 
of the absorber galaxies. Furthermore, more narrow-band imaging or spectroscopic 
studies in the near-infrared will give access to the H-$\alpha$ emission line 
from the $z \sim 2.5$ absorbers. A systematic comparison of the H-$\alpha$ and 
Ly-$\alpha$ 
data for the same absorbers can help to understand whether the emission lines in the galaxies in absorber fields 
are simply attenuated by dust, or whether these galaxies truly 
have low SFRs. It will thus be interesting to look at Ly-$\alpha$ emission 
from the objects 
reported to show strong H-$\alpha$ emission, and to look at H-$\alpha$ 
near the quasars for which we have limits on Ly-$\alpha$. Finally, more high-resolution narrow-band imaging and/or spectroscopic 
studies would be especially important to understand whether the star formation 
in the absorber galaxies is restricted to compact regions.

\acknowledgments

This paper is based on observations obtained with the Apache Point Observatory 
3.5-meter telescope, which is owned and operated by the 
Astrophysical Research Consortium. We thank Jonathan 
Brinkmann for assistance with the Fabry-Perot setup and calibration at the 
APO. We also thank Andrew Bunker for helpful discussion and Betty Stobie for providing and assisting with 
the IDP-3 package. Finally, we are grateful to an anonymous referee whose 
comments helped to improve this paper. VPK, DGT, and JM acknowledge support from the National 
Science Foundation grant AST-0206197 and from the University of South 
Carolina Research Foundation.

\clearpage

\begin{figure}
{\centerline{\hbox{\hsize=7.0truecm   
}}
\vskip 0.1in 
\centerline{\hbox{\hsize=7.0truecm 
}}
}
{\em {{\bf FIG. 1--} 
APO images of the field of Q0216+080. (a) $2.1 \arcmin  \times 2.1 \arcmin$ 
B-band image 
(top left); (b) $2.1 \arcmin  \times 2.1 \arcmin$ narrow-band image 
before continuum subtraction (top right); 
(c) $2.1 \arcmin  \times 2.1 \arcmin$ narrow-band image 
after continuum subtraction (bottom left); and (d) 
$1 \arcmin \times 1 \arcmin$ narrow-band image after continuum 
subtraction (bottom right). See attached gif images.} }
\end{figure} 
\clearpage

\begin{figure}
{\hskip 0.38in{\hbox{\hsize=7.0truecm   
}}
\vskip 0.1in 
\centerline{\hbox{\hsize=7.0truecm 
}}
}
{\em {{\bf FIG. 2--} Same as for Fig. 1, for the field of Q0636+680.    
} }
\end{figure} 
\clearpage

\begin{figure}
{\centerline{\hbox{\hsize=7.0truecm   
}}
\vskip 0.1in 
\centerline{\hbox{\hsize=7.0truecm 
}}
}
{\em {\baselineskip=8pt {\bf FIG. 3--} Same as for Fig. 1, for the field of Q0956+123.
} }
\end{figure} 
\clearpage

\begin{figure}
{\hskip 0.52in{\hbox{\hsize=6.5truecm   
}}
\vskip 0.1in 
\centerline{\hbox{\hsize=6.5truecm 
}}
}
{\em {{\bf FIG. 4--} Same as for Fig. 1, for the field of Q1209+093.    
} }
\end{figure} 
\clearpage

\begin{figure}
{\centerline{\hbox{\hsize=7.0truecm   
}}
\vskip 0.1in 
\centerline{\hbox{\hsize=7.0truecm 
}}
}
{\em {\baselineskip=8pt {\bf FIG. 5--} Same as for Fig. 1, for the field of Q1442+101.
} }
\end{figure} 
\clearpage

\begin{figure}
{\centerline{\hbox{\hsize=7.0truecm   
}}
\vskip 0.1in 
\centerline{\hbox{\hsize=7.0truecm 
}}
}
{\em {\baselineskip=8pt {\bf FIG. 6--} Same as for Fig. 1, for the field of Q2233+131.
} }
\end{figure} 
\clearpage

\begin{figure}
{\centerline{\hbox{\hsize=7.0truecm   
}}
\vskip 0.1in 
\centerline{\hbox{\hsize=7.0truecm 
}}
}
{\em {\baselineskip=8pt {\bf FIG. 7--} Same as for Fig. 1, for 
the calibration field of 53w002.
} }
\end{figure} 
\clearpage

\begin{figure}
\epsscale{0.3}
\centerline{\hbox{\hsize=13truecm
\epsfysize \hsize \epsfbox{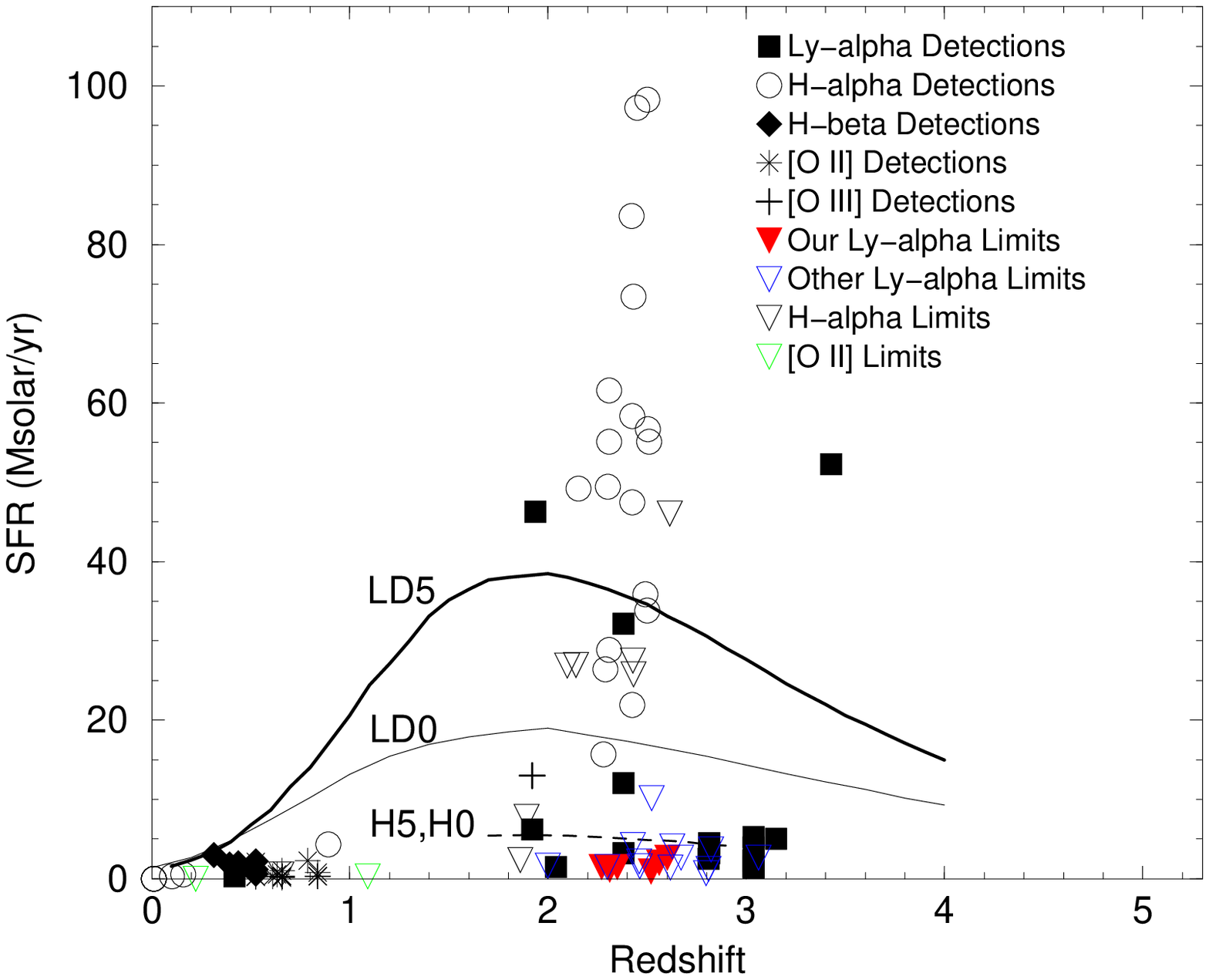}}}
{\em {\baselineskip=8pt { \bf FIG. 8--} Measurements of star formation rates (in $M_{\odot}$ yr$^{-1}$) 
for candidate objects in quasar 
absorber fields, from narrow-band imaging and spectroscopic 
searches for Ly-$\alpha$, H-$\alpha$, H-$\beta$, [O II] and [O III] emission 
lines. Data points are 
from our APO Ly-$\alpha$ survey (this work), our previous NICMOS H-$\alpha$ 
imaging (Kulkarni et al. 2000, 2001), and other literature 
(see text). Most of the higher values are based on candidate 
H-$\alpha$ emitters, often located far from the quasar lines of sight. The 
thick solid (upper) and thin solid (middle) curves show, respectively, 
the calculations of Bunker 
et al. (1999) for the predicted cross-section-weighted SFR in the large-disk 
scenario for $q_{0} = 0.5$ and $q_{0} = 0$. The dashed curve shows 
the calculations of Bunker et al. (1999) for the hierarchical scenario 
for $q_{0} = 0.5$ and $q_{0} = 0$. All of these curves are based on 
the closed-box global SFR models of Pei \& Fall (1995). }}
\end{figure}

%% If you are not including electonic art with your submission, you may
%% mark up your captions using the \figcaption command. See the
%% No more than seven \figcaption commands are allowed per page,
%% so if you have more than seven captions, insert a \clearpage

%% Tables should be submitted one per page, so put a \clearpage before

%% Two options are available to the author for producing tables:  the
%% deluxetable environment provided by the AASTeX package or the LaTeX
%% table environment.  Use of deluxetable is preferred.
%%

%% Three table samples follow, two marked up in the deluxetable environment,
%% one marked up as a LaTeX table.

%% In this first example, note that the \tabletypesize{}
%% command has been used to reduce the font size of the table.
%% We also use the \rotate command to rotate the table to
%% landscape orientation since it is very wide even at the
%% Note also that the \label command needs to be placed

%% This table also includes a table comment indicating that the full
%% version will be available in machine-readable format in the electronic
\clearpage

\begin{table}
\tabletypesize{\scriptsize}
\begin{center}
\caption{List of Targets\label{tbl-1}}
\begin{tabular}{llllllr}
\tableline\tableline
QSO & R.A.(J2000) & Dec.(J2000) & $z_{em}$ & $z_{abs}$ & Known Ions&
$W_{\rm{Ly-\alpha}}^{\rm{rest}}$ \\
&&&&&&({\AA})\\
\tableline
Q0216+080&02:18:57.3 &  +08:17:28.0&2.9957&2.2931&CIV,AlIII,AlII,&...$^\dagger$\\
&&&&&SiII,FeII; DLA\\
Q0636+680&06:42:04.3 &  +67:58:35.6& 3.1775&2.3112&CIV,CII,AlII&... \\
Q0956+123&09:58:52.3 &  +12:02:43.2& 3.306&2.3104&CIV,SiIV,AlII&1.42\\
Q1209+093&12:11:34.9 &  +09:02:22.8& 3.297&2.5822&CIV,AlII,FeII;&21.2\\
&&&&&DLA\\
Q1442+101&14:45:16.5 &  +09:58:36.2&  3.535&2.5632&OVI, CIV, SiIV, &1.40\\&&&&&CIII, SiIII, SiII, OI\\
Q2233+131&22:36:19.2&+13:26:20.0&3.2978&2.5537&CIV,CII,SiIV,SiIII,&5.35\\
&&&&&SiII,OI; sub-DLA\\
53w002&17:14:14.7&+50:15:29.7&...&...&...&...\\
(Calibration&...&...&...&...&...&...\\
Field)&...&...&...&...&...&...\\
\tableline
\end{tabular}

%% Any table notes must follow the \end{tabular} command.

\end{center}{\noindent $\dagger$: log $N_{\rm H I} =20.45$. }
\end{table}

\clearpage
\begin{deluxetable}{lcccccccc}
\tabletypesize{\scriptsize}
\tablecaption{Journal of Observations \label{tbl-2}}
\tablewidth{0pt}
\tablehead{
\colhead{Object} & 
\colhead{UT Date}  & \colhead{UT} & \colhead{sec$(z)$} &
\colhead{Etalon}     & \colhead{Filter}  &
\colhead{$\lambda_{\rm {cent}}$}   & \colhead{$t_{\rm{exp}}$(s)} & \colhead{Avg. FWHM 
({\AA})}
}
\startdata

%Q0216+080 &2000/10/26 &06:18:55.61 &1.117 &Blue &4050/160 &4002.2 &2400 & 14.69\\
%	  &2000/10/26&07:21:25.45&1.102&Blue&4050/160&4002.2&2400&\\
%	  ...&2000/10/26&08:08:18.24&1.142&Blue&4050/160&4002.2&2400&\\
%	  ...&2000/10/27&07:10:27.81&1.100&Blue&4050/160&4002&2400&\\
%	  ...&2000/10/27&07:57:04.58&1.133&Blue&4050/160&4002&2400&\\
%	  ...&2000/10/27&08:38:42.29&1.203&Blue&4050/160&4002&2400&\\	  
%	  ...&2001/11/09&06:43:51.17&1.111&Blue&4050/160&4002.2&2400&\\
Q0216+080 &2001/11/10&05:55:28.87&1.111&Blue&4050/160&4002.2&2400&14.69\\
	  ...&2001/11/10&06:36:18.15&1.111&Blue&4050/160&4002.2&1200&\\
	  ...&2001/11/11&05:42:48.66&1.102&Blue&4050/160&4002.2&2400&\\
	  ...&2001/11/11&06:52:30.75&1.125&Blue&4050/160&4002.2&1660&\\
%	  ...&2001/11/12&04:36:20.36&1.164&Blue&4050/160&4001.3&2400&\\
	  ...&2001/11/12&05:20:37.43&1.111&Blue&4050/160&4001.3&2400&\\
	  ...&2001/11/12&06:05:54.94&1.100&Blue&4050/160&4001.3&2400&\\
%	  ...&2001/12/15&02:08:01.90&1.200&Blue&4050/160&4002.2&2400&\\
	  ...&2001/12/15&03:17:36.77&1.107&Blue&4050/160&4002.2&2400&\\
	  ...&2001/12/15&04:00:38.97&1.101&Blue&4050/160&4002.2&2400&\\
	  ...&2001/12/15&04:50:23.34&1.139&Blue&4050/160&4002.2&2400&\\
	  ...&2001/12/15&05:32:31.49&1.216&Blue&4050/160&4002.2&2400&\\
	  ...&2002/11/03&06:59:28.92 & 1.106 & Vis-B& 4050/160&4008.0 & 2400& \\
	  ...&2002/11/03&07:52:40.87&1.163&Vis-B&4050/160&4008.0 & 2400 &\\
	  ...&2002/11/08&07:42:27.53&1.180&Vis-B&4050/160&4008.0& 1800 &\\
%	  ...&2002/11/08&08:50:26.30&1.387&Vis-B&4050/160&4008.0& 1800& \\
&&&&&&&\\  
          ...&2000/10/26&05:23:00.92&1.201&...&B&4400&480&\\
	  ...&2000/10/26&05:32:58.88&1.180&...&B&4400&480&\\
	  ...&2000/10/26&05:42:02.58&1.164&...&B&4400&480&\\
	  ...&2000/10/26&05:51:06.28&1.150&...&B&4400&480&\\
	  ...&2000/10/26&06:00:28.58&1.137&...&B&4400&480&\\
...&2001/12/15&01:21:29.11&1.338&...&B&4400&300&\\
...&2001/12/15&01:27:45.25&1.314&...&B&4400&300&\\
...&2001/12/15&01:34:03.80&1.292&...&B&4400&300&\\
...&2001/12/15&01:40:22.75&1.272&...&B&4400&300&\\
...&2001/12/15&01:46:35.10&1.254&...&B&4400&300&\\
&&&&&&&\\
\tableline
&&&&&&&\\
Q0636+680 &  2000/10/26&10:51:55.81&1.229&Blue&4050/160&4024&2400&14.66\\
...&2000/10/26&11:33:12.48&1.224&Blue&4050/160&4024&2400&\\
%...&2000/10/27&09:40:25.10&1.270&Blue&4050/160&4024&2400&\\
...&2001/02/22&03:20:22.82&1.225&Blue&4050/160&4024.3&2400&\\
...&2001/02/22&04:18:32.24&1.232&Blue&4050/160&4024.3&2400&\\
...&2001/02/22&05:06:11.28&1.260&Blue&4050/160&4024.3&2400&\\
...&2001/12/15&07:15:20.01&1.236&Blue&4050/160&4024.2&2400&\\
...&2001/12/15&07:56:38.23&1.224&Blue&4050/160&4024.2&2400&\\
...&2001/12/15&08:39:13.21&1.228&Blue&4050/160&4024.2&2400&\\
...&2001/12/15&09:27:39.32&1.252&Blue&4050/160&4024.2&2400&\\
%...&2002/03/09&03:55:01.38&1.250&Blue&4050/160&4024.2&2400&\\
%...&2002/03/09&05:17:25.68&1.344&Blue&4050/160&4024.2&2400&\\
%...&2002/03/10&04:02:13.07&1.259&Blue&4050/160&4024.2&2400&\\
...&2002/03/12&02:56:26.66&1.228&Blue&4050/160&4024.2&2400&\\
...&2002/03/12&03:38:47.01&1.247&Blue&4050/160&4024.2&2400&\\
%...&2002/11/03&08:56:12.99&1.290&Vis-B&4050/160&4030.0&2400&\\
%...&2002/11/03&09:38:12.51&1.251&Vis-B&4050/160&4030.0&2400&\\
...&2002/11/03&10:45:35.05&1.224&Vis-B&4050/160&4030.0&2400&\\
...&2002/11/03&11:29:28.69&1.229&Vis-B&4050/160&4030.0&2400&\\
...&2002/11/05&09:44:12.32&1.242&Vis-B&4050/160&4030.0&2400&\\
%...&2002/11/06&09:42:37.47&1.241&Vis-B&4050/160&4030.0&2400&\\
&&&&&&&\\
...&2000/10/26&10:07:53.96&1.251&...&B&4400&480&\\
...&2000/10/26&10:17:24.49&1.244&...&B&4400&480&\\
...&2000/10/26&10:26:39.20&1.239&...&B&4400&480&\\
...&2000/10/26&10:35:43.11&1.235&...&B&4400&480&\\
...&2001/02/25&05:16:15.67&1.281&...&B&4400&300&\\
...&2001/02/25&05:22:41.63&1.288&...&B&4400&300&\\
...&2001/02/25&05:29:27.60&1.296&...&B&4400&300&\\
&&&&&&&\\
\tableline
&&&&&&&\\
Q0956+123&2001/02/22&06:23:39.98&1.077& Blue&4050/160& 4023.3&2400&14.79\\
...&2001/02/22&07:24:08.73&1.078&Blue&4050/160& 4023.3&2400&\\
...&2001/02/22&08:16:52.25&1.135&Blue&4050/160& 4023.3&2400&\\
...&2001/02/22&09:24:17.19&1.312&Blue&4050/160& 4023.3&2400&\\
%...&2001/02/25&07:53:52.54&1.118&Blue&4050/160&4023.3&2400&\\
...&2001/02/25&08:38:17.40&1.204&Blue&4050/160&4023.4&2400&\\
%...&2001/05/21&03:23:18.17&1.260&Blue&4050/160&4023.2&2400&\\
...&2001/05/22&03:18:47.08&1.258&Blue&4050/160&4023.3&2400&\\
%...&2001/12/15&11:08:47.14&1.072&Blue&4050/160&4023.3&2400&\\
%...&2002/03/10&05:45:06.19&1.070&Blue&4050/160&4023.2&2400&\\
%...&2002/03/10&06:43:33.25&1.094&Blue&4050/160&4023.2&2400&\\
%...&2002/03/10&07:45:15.89&1.198&Blue&4050/160&4023.2&\\
...&2002/03/12&05:13:05.87&1.078&Blue&4050/160&4023.2&2400&\\
...&2002/03/12&05:55:28.03&1.071&Blue&4050/160&4023.2&2400&\\
...&2002/03/12&06:44:56.70&1.104&Blue&4050/160&4023.2&2400&\\
%...&2002/05/11&03:14:48.09&1.142&Blue&4050/160&4023.2&2400&\\
%...&2002/05/13&03:11:25.90&1.150&Blue&4050/160&4023.2&2400&\\
&&&&&&&\\
...&2001/02/25&06:10:45.09&1.078&...&B&4400&300&\\
...&2001/02/25&06:17:40.16&1.074&...&B&4400&300&\\
...&2001/02/25&06:24:01.83&1.072&...&B&4400&300&\\
...&2001/02/25&06:30:58.02&1.070&...&B&4400&300&\\
...&2001/02/25&06:37:32.38&1.069&...&B&4400&300&\\
...&2001/02/25&06:49:41.66&1.069&...&B&4400&300&\\
...&2001/02/25&07:06:17.69&1.074&...&B&4400&300&\\
...&2001/02/25&07:12:54.06&1.078&...&B&4400&300&\\
&&&&&&&\\
\tableline
%&&&&&&&\\
%Q1202-0054&2003/03/31&06:16:47.73&1.205&Vis-B&4400/135&4448.6&2400&\\
%...&2003/03/31&06:58:32.83&1.210&Vis-B&4400/135&4448.6&2400&\\
%...&2003/03/31&07:47:21.86&1.270&Vis-B&4400/135&4448.6&2400&\\
%
%&&&&&&&\\
%...&2003/03/31&05:44:33.34&1.229&...&B&4400&600&\\
%...&2003/03/31&05:56:34.81&1.217&...&B&4400&600&\\
%...&2003/03/31&08:31:29.90&1.387&...&B&4400&600&\\
%&&&&&&&\\
%\tableline
&&&&&&&\\
Q1209+093 &2001/05/20&04:42:28.87&1.154&Blue&4300/135&4353.6&2400&7.01\\
...&2001/05/20&05:28:37.23&1.260&Blue& 4300/135& 4353.6&2400&\\
%...&2001/05/21&04:33:25.44&1.146&Blue& 4300/135&4353.5&2400&\\
%...&2001/05/21&05:15:10.77&1.233&Blue& 4300/135&4353.5&2400&\\
...&2001/05/22&05:11:26.45&1.233&Blue& 4300/135&4353.5&2400&\\
...&2001/05/23&03:41:53.18&1.101&Blue& 4300/135&4353.7&2400&\\
...&2001/05/23&04:24:29.14&1.145&Blue& 4300/135&4353.6&2400&\\
%...&2002/03/09&07:04:37.05&1.131&Blue& 4300/135&4353.5&2400&\\
%...&2002/03/09&08:34:54.56&1.100&Blue& 4300/135&4353.5&2400&\\
%...&2002/03/09&09:18:01.28&1.141&Blue& 4300/135&4353.5&2400&\\
...&2002/03/11&08:52:16.27&1.119&Blue& 4300/135&4353.5&2400&\\
%...&2002/03/12&08:24:36.65&1.100&Blue&5700/135&4353.5&&\\
...&2002/03/12&08:25:38.11&1.101&Blue&4300/135&4353.5&2400&\\
...&2002/03/12&09:17:40.82&1.159&Blue&4300/135&4353.5&2400&\\
%...&2002/05/11&05:25:20.72&1.166&Blue&4300/135&4353.5&2400&\\
%...&2002/05/13&04:32:46.00&1.108&Blue&4300/135&4353.5&2400&\\
%...&2002/05/13&05:14:07.50&1.160&Blue&4300/135&4353.5&2400&\\
&&&&&&&\\
...&2001/05/23&05:22:34.56&1.277&...&B&4400&300&\\
...&2001/05/23&05:37:57.81&1.332&...&B&4400&300&\\
...&2001/05/23&05:44:12.36&1.357&...&B&4400&300&\\
...&2001/05/23&05:50:36.43&1.384&...&B&4400&300&\\
...&2001/05/23&05:56:55.87&1.413&...&B&4400&300&\\
...&2001/05/23&06:03:07.01&1.444&...&B&4400&300&\\
...&2001/05/23&06:09:17.35&1.477&...&B&4400&300&\\
...&2001/05/23&06:15:38.11&1.514&...&B&4400&300&\\

\tableline
&&&&&&&\\

%Q1442+101 &...&2001/02/25&09:47:54.47&1.183&...&B&4400&120\\
%...&2001/02/25&09:55:24.48&1.168&...&B&4400& 120\\
%Q1442+101&2001/05/20&06:47:53.02&1.110&Blue&4300/135&4330.4&2400&6.10\\
Q1442+101&2001/05/20&07:29:32.74&1.170&Blue&4300/135&4330.5&2400&6.10\\
...&2001/05/20&08:17:28.39&1.299&Blue&4300/135&4330.4&2400&\\
...&2001/05/21&06:37:53.56&1.104&Blue&4300/135&4330.6&2400&\\
...&2001/05/21&07:27:17.70&1.174&Blue&4300/135&4330.5&2400&\\
...&2001/05/21&08:09:09.42&1.284&Blue&4300/135&4330.4&2400&\\
...&2001/05/22&06:15:29.60&1.091&Blue&4300/135&4330.5&2400&\\
...&2001/05/22&06:57:37.14&1.130&Blue&4300/135&4330.5&2400&\\
%...&2001/05/22&07:47:11.69&1.230&Blue&4300/135&4330.3&2400&\\
...&2001/05/23&06:51:26.89&1.127&Blue&4300/135&4330.5&2400&\\
...&2001/05/23&07:33:07.00&1.204&Blue&4300/135&4330.5&2400&\\
%...&2002/03/09&10:31:51.20&1.086&Blue&4300/135&4330.5&2400&\\
%...&2002/03/09&11:15:29.22&1.096&Blue&4300/135&4330.5&2400&\\
%...&2002/03/11&10:08:07.28&1.091&Blue&4300/135&4330.5&2400&\\
%...&2002/03/11&10:50:07.23&1.088&Blue&4300/135&4330.5&2400&\\
%...&2002/03/12&10:24:42.14&1.085&Blue&4300/135&4330.5&2400&\\
%...&2002/03/12&11:08:10.57&1.099&Blue&4300/135&4330.5&2400&\\
%...&2002/03/12&12:19:12.73&1.209&Blue&4300/135&4330.5&900&\\
%...&2002/05/11&06:36:31.67&1.085&Blue&4300/135&4330.5&2400&\\
%...&2002/05/11&07:17:06.92&1.103&Blue&4300/135&4330.5&2400&\\
%...&2002/05/11&07:24:34.93&1.110&Blue&4300/135&4330.5&2400&\\
%...&2002/05/11&08:09:28.03&1.177&Blue&4300/135&4330.5&2400&\\
%...&2002/05/14&07:06:39.99&1.104&Blue&4300/135&4330.5&2400&\\
%...&2002/05/14&07:48:02.69&1.159&Blue&4300/135&4330.5&2400&\\
&&&&&&&\\
...&2001/02/25&10:03:28.53&1.153&...&B&4400&300&\\
...&2001/02/25&10:10:44.33&1.141&...&B&4400& 300&\\
...&2001/02/25&10:18:19.75&1.130&...&B&4400&300&\\
...&2001/02/25&10:26:38.82&1.120&...&B&4400&300&\\
%...&2001/02/25&10:46:06.10&1.101&...&B&4400&60&\\
%...&2001/02/25&10:51:00.57&1.097&...&B&4400&60&\\
%...&2001/02/25&10:58:10.17&1.093&...&B&4400& 60&\\
%...&2001/02/25&11:03:33.67&1.090&...&B&4400&60&\\
%...&2001/02/25&11:16:04.76&1.086&...&B&4400&60&\\
...&2001/02/25&11:21:21.65&1.085&...&B&4400&300&\\
...&2001/02/25&11:32:29.47&1.085&...&B&4400&300&\\
...&2001/02/25&11:41:09.15&1.087&...&B&4400&300&\\
...&2001/02/25&11:48:41.57&1.089&...&B&4400&300&\\
...&2001/02/25&11:55:50.50&1.0938&...&B&4400&300&\\
...&2001/05/23&08:21:14.63&1.359&...&B&4400&300&\\
...&2001/05/23&08:27:35.43&1.387&...&B&4400&300&\\
...&2001/05/23&08:33:59.98&1.418&...&B&4400&300&\\
&&&&&&&\\
\tableline
&&&&&&&\\
%Q2233+131 & 2000/10/25&03:18:23.49&1.060&Blue&4300/135&4320&2400&5.59\\
%...&2000/10/25&04:09:10.69&1.078&Blue&4300/135&4320&2400&\\
Q2233+131&2000/10/26&03:45:44.77&1.066&Blue&4300/135&4319&2400&5.59\\
...&2000/10/27&01:38:14.59&1.150&Blue&4300/135&4314.3&2400&\\
...&2000/10/27&02:41:43.61&1.070&Blue&4300/135&4317.0&2400&\\
...&2000/10/27&03:32:30.68&1.062&Blue&4300/135&4318&2400&\\
...&2000/10/27&04:20:34.25&1.097&Blue&4300/135&4318&2400&\\
...&2000/10/27&05:12:21.30&1.190&Blue&4300/135&4318.5&2400&\\
...&2000/10/27&06:11:24.96&1.398&Blue&4300/135&4319&2400&\\
...&2001/11/10&02:29:02.38&1.060&Blue&4313/116&4318.9&2400&\\
...&2001/11/10&03:10:40.90&1.081&Blue&4313/116&4318.9&2400&\\
...&2001/11/10&03:52:49.45&1.136&Blue&4313/116&4318.9&2400&\\
...&2001/11/10&04:34:24.56&1.233&Blue&4313/116&4318.9&2400&\\
...&2001/11/11&02:01:10.21&1.062&Blue&4313/116&4318.9&2400&\\
...&2001/11/11&02:45:46.09&1.066&Blue&4313/116&4318.9&2400&\\
%...&2001/11/11&04:36:48.47&1.253&Blue&4313/116&4318.9&2400&\\
...&2001/11/12&02:18:11.07&1.060&Blue&4313/116&4318.9&2400&\\
...&2001/11/12&03:01:43.69&1.080&Blue&4313/116&4318.9&2400&\\
%...&2002/11/04&01:51:25.80& 1.087&Vis-B&4300/135&4325.0&2400&\\
...&2002/11/04&03:08:18.76& 1.064&Vis-B&4300/135&4325.0&1740&\\
...&2002/11/06&01:29:15.00&1.104&Vis-B&4300/135&4325.0&2400&\\
...&2002/11/06&02:31:17.24&1.060&Vis-B&4300/135&4325.0&800&\\
...&2002/11/06&05:02:00.19&1.268&Vis-B&4300/135&4325.0&2400&\\
&&&&&&&\\
...&2000/10/26&02:42:34.68&1.072&...&B&4400&480&\\
...&2000/10/26&02:53:33.87&1.066&...&B&4400&480&\\
...&2000/10/26&03:06:41.39&1.061&...&B&4400&480&\\
...&2000/10/26&03:17:01.37&1.060&...&B&4400&480&\\
...&2000/10/26&03:28:07.59&1.060&...&B&4400&480&\\
...&2001/11/11&03:54:43.13&1.147&...&B&4400&2400&\\
...&2000/10/25&05:26:24.96&1.205&...&B&4400&1200&\\
&&&&&&&\\
\tableline
&&&&&&&\\
53w002&2001/05/20&09:45:49.27&1.086&Blue&4155/160&4128.4&2400&8.31\\
...&2001/05/21&09:34:32.16&1.079&Blue&4155/160&4128.5&2400&\\
...&2001/05/21&10:15:54.17&1.126&Blue&4155/160&4128.6&2400&\\
...&2001/05/22&09:08:42.61&1.063&Blue&4155/160&4128.3&2400&\\
...&2001/05/23&09:31:39.77&1.084&Blue&4155/160&4128.3&2400&\\
&&&&&&&\\

...&2001/05/22&09:57:53.15&1.108&...& B&4400&300&\\
...&2001/05/22&10:04:47.73&1.117&...& B&4400&300&\\
...&2001/05/22&10:10:54.66&1.125&...& B&4400&300&\\
...&2001/05/22&10:17:17.82&1.134&...& B&4400&300&\\
...&2001/05/22&10:23:55.19&1.145&...& B&4400&300&\\
...&2001/05/22&10:34:30.57&1.163&...& B&4400&300&\\
...&2001/05/22&10:41:24.96&1.176&...& B&4400&300&\\
...&2001/05/22&10:52:14.56&1.197&...& B&4400&300&\\
...&2001/05/23&08:44:33.56&1.054&...& B&4400&300&\\
...&2001/05/23&08:51:20.34&1.057&...& B&4400&300&\\
...&2001/05/23&08:57:47.50&1.060&...& B&4400&300&\\
...&2001/05/23&09:05:17.92&1.064&...& B&4400&300&\\

\tableline
&&&&&&&\\
 \enddata
%comment: May 14: checked May 01, Dec 01; Previously had checked Feb 01, Nov01 
% for inclusion in above table. May 15:checked mar 02. May 20: checked may 02.

%% Text for table notes should follow after the \enddata but before
%% the \end{deluxetable}. Make sure there is at least one \tablenotemark
%% in the table for each \tablenotetext.
%
%\tablenotetext{a}{Sample footnote for table~\ref{tbl-1} that was generated
%with the deluxetable environment}
%\tablenotetext{b}{Another sample footnote for table~\ref{tbl-1}}
%
%\tablecomments{Occasionally, authors wish to append a short
%paragraph of explanatory notes that pertain to the entire table, but
%%which are different than the caption.  Such notes should be placed in
%a {\tt tablecomments} command like this.}

\end{deluxetable}
\clearpage

\begin{table}
\tabletypesize{\scriptsize}
\begin{center}
\caption{Lyman-$\alpha$ Point source flux sensitivities and Constraints on Star Formation Rates\label{tbl-3}}
\begin{tabular}{llcc}
\tableline\tableline
QSO & $z_{abs}$ & $f_{{\rm Ly}-\alpha} ^{\dagger}$(erg s$^{-1}$ cm$^{-2}$))&
SFR ($M_{\odot}$ yr$^{-1}$)$^{\dagger \dagger}$ \\
\tableline
Q0216+080&2.2931&$<4.0 \times 10^{-17}$&$ < 1.5$\\
Q0636+680&2.3112&$<3.2 \times 10^{-17}$&$< 1.2$ \\
Q0956+123&2.3104&$<3.9 \times 10^{-17}$&$ < 1.5$\\
Q1209+093&2.5822&$<5.4 \times 10^{-17}$&$< 2.7$\\
Q1442+101&2.5632&$<4.2 \times 10^{-17}$&$< 2.0$\\
Q2233+131&2.5537&$<1.9 \times 10^{-17}$&$< 0.9$\\
\tableline
\end{tabular}
\end{center}
{\noindent $\dagger$: 3 $\sigma$ upper limits on Lyman-$\alpha$ flux for a 
2 pixel x 2 pixel region, corresponding to a physical size of $6.0 h^{-1}$ kpc 
x 6.0 kpc at $z = 2.4$; 
$\dagger \dagger$: 3 $\sigma$ upper limits on SFR. }
%
%% Any table notes must follow the \end{tabular} command.
%
\end{table}
\clearpage
\end{document}